\documentclass[12pt,aps,prb,preprint]{revtex4}   

\usepackage{amsmath}    
\usepackage{graphicx}   

\begin{document}

\title{The concepts of work and heat and the first and second laws of thermodynamics}
\author{Rodrigo de Abreu}
 \affiliation{Departamento de F\'{\i}sica, Instituto Superior T\'ecnico, Universidade T\'ecnica de Lisboa, 1049-001 Lisboa, Portugal}
\author{Vasco Guerra}
 \affiliation{Departamento de F\'{\i}sica, Instituto Superior T\'ecnico, Universidade T\'ecnica de Lisboa, 1049-001 Lisboa, Portugal}
 \email{vguerra@ist.utl.pt}   
\date{\today}

\begin{abstract}
A simple and effective approach to thermodynamics is suggested, which
solves the major difficulties in the traditional presentation of the subject. 
The internal energy is introduced from the behavior of deformable bodies,  whereas the importance of 
keeping in mind the microscopic picture is emphasized.
A straightforward model is used to show that the internal energy depends on the volume and entropy, from where
the relationship between mechanics and thermodynamics is immediate, mechanics corresponding to
isentropic thermodynamics. The questions of evolution to equilibrium and irreversibility are
studied under the light of  the action of the ``dynamic force,'' which has a dissipative character.
The present formulation leads to a discussion and clarification of the physical meaning of various thermodynamic quantities, such as pressure, 
temperature, work and heat.
The adiabatic piston problem is analyzed, as a paradigmatic case where the notions of ``adiabatic''
and ``heat transfer'' are often ill-defined.
\end{abstract}

\maketitle

\section{Introduction}

Thermodynamics has been, still is and will remain to be one of the key disciplines of physics
and chemistry. However, in its traditional form it has recurrently brought intellectual resistance.
For instance, there is a famous story about Arnold Sommerfeld and thermodynamics that is cited in several sources\cite{Angetal1967,Fink2009}
and spreads on the internet.
When Sommerfeld was asked why he had never written  a book on thermodynamics, since he had written books on nearly every topic of physics,
 his humorous answer was
``Thermodynamics is a funny subject. The first time you go through it, you don't understand it at all. The second time you go through it, you think you understand it, except for one or two points. The third time you go through it, you know you don't understand it, but by that time you are so used to that subject, it doesn't bother you anymore.'' 
Actually Sommerfeld died in 1951 and at the time he was writing a book on thermodynamics and statistical mechanics, which was published after 
his death.\cite{Sommerfeld1954}
The interesting fact in this story is that thermodynamics, despite being well established for many years and dealing with apparently simple and intuitive concepts, 
nevertheless remains quite difficult and subtle. 

We believe the difficulties with thermodynamics arise essentially from two main factors. 
The first one is the traditional development of the concepts and the formalism using the
variable \textit{temperature}. As it is well-known, the natural variables associated
with the internal energy $U$ are the \textit{volume} and the \textit{entropy}, which allow to Legendre transform
$U$ into the other thermodynamic potentials and flank it on the 
``thermodynamic square.''\cite{Huang1987} Temperature and pressure can then be readily obtained from Maxwell's relations.
We consider that a formulation of thermodynamics
made considering from the beginning $U=U(S,V)$ contributes to a neat mathematical
derivation and an easier understanding of the meaning of the different symbols.
An early introduction of the entropy was also proposed by Herbert Callen.\cite{Callen1985}

A second difficulty comes from the fact that thermodynamic quantities such as temperature and pressure are defined in equilibrium. 
The application of the formalism to non-equilibrium processes 
makes several variables and notions lose their intuitive physical meaning and the mere use of the words denoting these variables may induce in error.
The difference between the static and dynamic pressures is crucial and is thoroughly discussed in this paper.
Another paradigmatic example is the ``adiabatic piston'' problem briefly reviewed in Section~\ref{adiabaticpiston}, where the 
word ``adiabatic'' is itself problematic. 

Richard Feynman, in the very first chapter of his ``Lectures on Physics,''\cite{Feyetal1979} addresses his students in
the following way: ``If, in some cataclysm, all of scientific knowledge were to be destroyed, and only one sentence
passed to the next generations of creatures, what sentence would contain the most information in the fewest words?
I believe it is the \textit{atomic hypothesis} (or the atomic \textit{fact}, or whatever you wish to call it) that
\textit{all things are made of atoms -- little particles moving around in perpetual motion, attracting each other when they are a little distance apart, but
repelling upon being squeezed into one another}.'' We like to recall this remarkable start of Feynman's ``Lectures.'' 
Its pertinence
to the present paper resides in the fact that
one way to minimize the difficulties surrounding thermodynamics is to always keep in mind the microscopic
mechanisms in which it is founded.  

In this paper we want to suggest a way to avoid most of the traps frequently encountered in the presentation of thermodynamics.
To this purpose, we put forward a development where we try to keep the ideas very simple, by maintaining a natural and strong connection
with mechanics and the microscopic view. We start by making the link between mechanical energy and the internal energy of deformable
bodies in Section~\ref{WorkEnergy}. In Section~\ref{base} we use a simple and clear ``base model''
to shown that the internal energy cannot be a function only of the deformation variable.
Entropy is introduced and the relationship with mechanics is immediate: mechanics is isentropic thermodynamics. The second law and the
questions of evolution to  equilibrium and irreversibility are also addressed in Section~\ref{base}, which finishes with the presentation of the formalism.
Heat and the first law of thermodynamics are introduced in Section~\ref{Generalizations}, from generalizations to the base model. The 
important case of the adiabatic piston
is analyzed in this chapter as well, as an illustration of the advantages of the proposed approach. Further extensions of the model, leading
to the detection and clarification of additional nuances, are point out and briefly debated.
The main findings are reviewed and discussed in Section~\ref{discussion}, which concludes the paper.

\section{Work and energy: mechanics and thermodynamics}\label{WorkEnergy}

As it is well-known, the concepts of \textit{work} and \textit{energy} were developed empirically, associated first with the notions of \textit{force}, 
\textit{gravitational potential energy} and \textit{kinetic energy}. The complexity of the action of muscles was substituted by a simplified description
of a static force, which can be measured, for instance, by a dynamometer. This lead to the notions of \textit{weight} as a force, as well as of the corresponding 
opposing force present in equilibrium.
The concepts of work of the weight and work of the force opposing weight appear naturally from here. The idea of gravitational potential energy arises
from the design and construction of weight lifting machines. 

All these intuitive concepts are very fundamental and part of any introductory
study of physics. Among many excellent textbooks available, we would like to recall the brilliant presentation of the subject by Feynman,\cite{Feyetal1979}
who introduces it following the route just described.
Another hypothesis, which he immediately advances and was early noted by Leonardo da Vinci,\cite{Leonardo}
is that ``there is no such thing as perpetual motion with weight lifting machines.'' Feynman moves on to warn that ``in fact, that there is no perpetual motion at all
is a general statement of the law of conservation of energy.''

Newton's laws allow a general description of motion. The identification of the variation of the kinetic energy as the work done by the resultant or net force 
applied to a particle of a certain mass is one of the results of the theory. Putting together gravitational potential energy and kinetic energy we have the conservation of
mechanical energy, valid for a system moving only under the effect of the gravitational force.

\begin{figure}[h]
\begin{center}
\scalebox{0.15}{\includegraphics{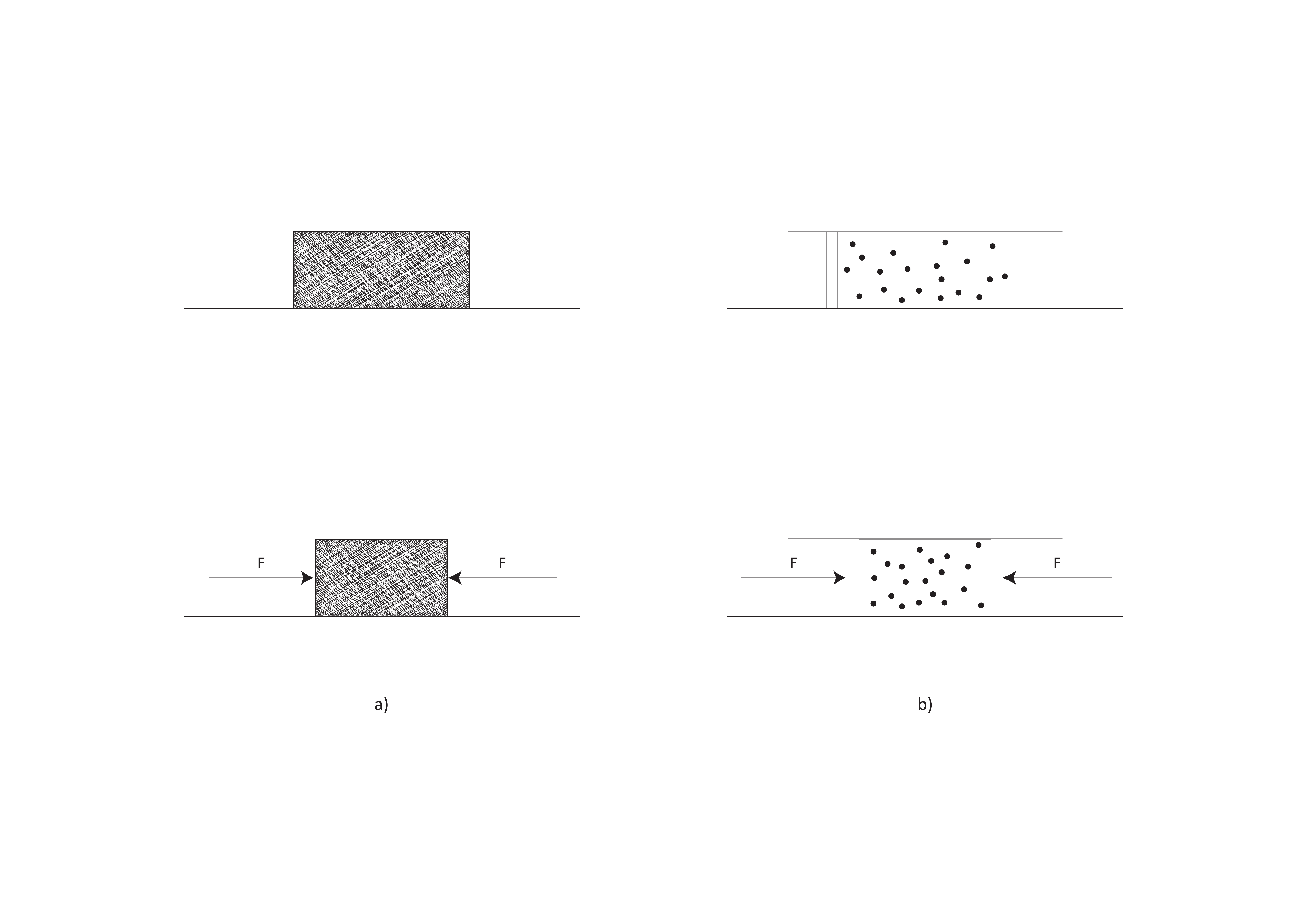}}
\caption{\label{deform}A body deforms under the action of two forces of
equal magnitude: a) macroscopic picture; b) microscopic picture, the body
being an ideal gas in a container with movable walls.}
\end{center}
\end{figure}

The generalization of these ideas to extended deformable bodies is at the origin of the notion of \textit{internal energy}. A very simple example is depicted in 
 Figure~\ref{deform}a). A pair of forces of the same magnitude and opposite
 directions acts on a body, deforming it. Clearly, there is work
 done by the forces, but the translation kinetic energy and the
gravitational potential energy of the body remain unchanged. 
 This straightforward example illustrates as well the power of the microscopic
 view as a model for thought.
The most simple microscopic picture is to imagine the
 body as an ideal gas in a container with movable walls, Figure~\ref{deform}b).
In this case the work done is converted in kinetic 
energy of the constituents of the gas, which is in fact its internal energy.

With further maturation, this extension of the concepts of mechanics
to deformable bodies is at the genesis of thermodynamics. 
Historically, however, 
thermodynamics  evolved in a rather autonomous way. Its development was based on the concepts of heat and temperature, with properties apparently foreign to
the phenomenology described by mechanics. These two new concepts were not easily encompassed by the formalism of mechanics, although there have always been
``kinetic'' conceptualizations of the thermodynamic quantities.\cite{Brush1986}

The application of Newtonian physics to engineering, military engineering being of particular importance, forced mechanics to face the question 
of \textit{friction}, which is always present in practice. Friction was -- and still is --  treated as a phenomenon which is not an essential part of mechanics.
One recurrent statement revealing the attempt to isolate mechanics from the phenomenology of friction is the affirmation of the reversibility of the equations
of mechanics. 

Friction brings to light that mechanics and thermodynamics are one and the same subject.
The works of Maxwell and Boltzmann culminate an amazing effort of many authors 
to conciliate mechanics with thermodynamics,
addressing the question of evolution to equilibrium.
Boltzmann, in particular, proposing a microscopic interpretation of entropy,\cite{Boltzmann1872,Boltzmann1877}
opens the door for a reevaluation of the meaning of several concepts, still problematic in the basic formulation of thermodynamics 
nowadays.\cite{Besson2003,Minguez2005}
Let us pass through the door and travel this road in the remaining of this paper.

\section{Energy and entropy: the base model}\label{base}

Thermodynamics is established for a long time and there
is a relatively consensual view on the way to present it. Most concepts, such
as work, temperature and heat, are usually
introduced from the very beginning and in a very intuitive way. 
Despite these facts, various difficulties arise in the interpretation
of some fundamental quantities -- heat and work on the first line -- even in quite 
simple situations. The subtleness of these seemingly simple and
unambiguous notions,  
lead to a permanent search for consistency in the definitions of work
and heat in the formulation of the first and second laws of thermodynamics,
still very active nowadays.
\cite{GC1987,Barrow1988,GC2005,GC2007,Canagaratna2008,Bizarro2008,Bizarro2010,Abreu1996,Abreu2002b,Gislason2010, AG2011}

We suggest an early introduction of the variable entropy, postponing
the appearance of the quantities temperature and heat.
This approach allows a formulation of thermodynamics that uses from the beginning
the natural variables associated with the internal energy, which are the volume and the
entropy. Moreover, all remaining quantities are introduced in a very general and clear
way, which, we believe, helps preventing the misunderstanding and errors that spread even
in the scientific literature.
Finally, this procedure immediately provides an easily identifiable connection between
mechanics and thermodynamics. 
 
\subsection{The need for the variable entropy}\label{needentropy}
\begin{figure}
\begin{center}
\scalebox{0.20}{\includegraphics{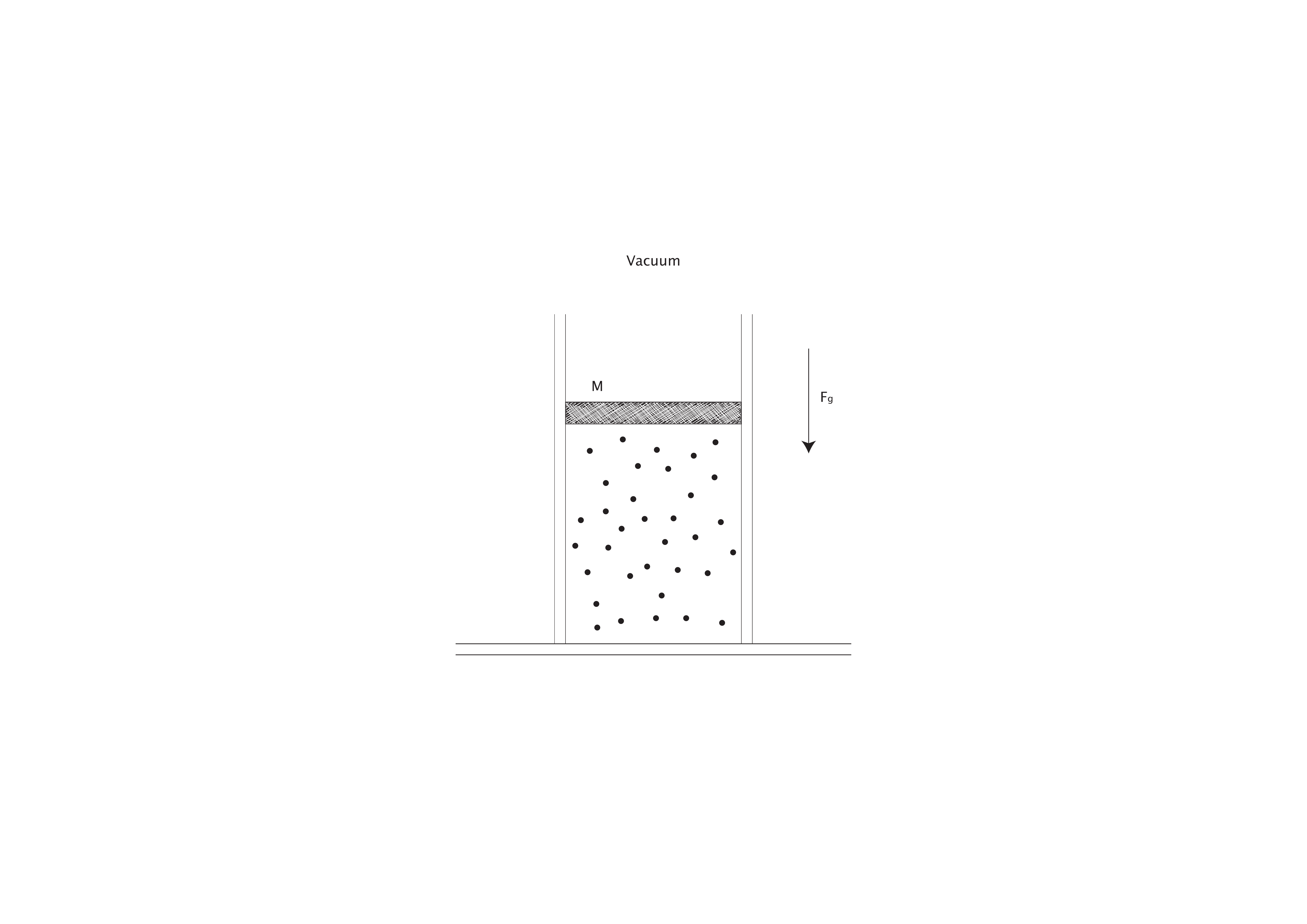}}
\caption{\label{basemodel}The base model: an ideal gas is contained in a cylinder with
a frictionless movable piston of mass $M$ on the top.}
\end{center}
\end{figure}

Thermodynamics can be introduced with generality from a clear, uncomplicated, 
``base model,''
corresponding to one of the most typical 
systems in thermodynamics: a classical ideal gas in a container with a
movable piston on top, under the action of gravity, as shown in Figure~\ref{basemodel}.
The system is 
surrounded by vacuum, so that there is no external atmospheric pressure 
on the piston. There is no friction between the piston and the container walls.
Furthermore, it is assumed that the piston and the container walls do not have any
internal structure, so that all collisions between the gas particles and the piston or the containing walls
are perfectly elastic.
This system can be studied starting only with the notions of weight, work done by the weight,
gravitational potential energy and internal energy. Note that a similar system
was used by Kivelson and Oppenheim to discuss the concept of 
work in irreversible processes.\cite{KO1966} 

Let us assume that the gas initially occupies a volume $V_1$ and exerts a pressure
$P_1$ on the container walls. The piston is held fixed in its place. Its total mass, $M$, 
is such that the pressure exerted by the piston as soon as it is released, 
\begin{equation}\label{Pe}
P_e=\frac{Mg}{A}\ , 
\end{equation}
is larger than $P_1$, $A$ denoting the area of the base of the
piston.

The final equilibrium position of the piston can be determined if we know the
dependence of the gas pressure with the gas internal energy. For a classical monoatomic ideal
gas, from the microscopic picture and the calculation of the
average momentum transfer per unit time it is easy to show that this relation is 
\begin{equation}\label{pressure}
P=\frac{2}{3}\frac{U}{V} = \alpha\frac{U}{V}
\end{equation}
(notice that in fact all
results can be obtained assuming the relation $P=\alpha(U/V)$, where $\alpha$
is a constant parameter dependent on the gas, which describes also the cases
of diatomic and polyatomic ideal gases, as well as the relativistic gas of photons with $\alpha=1/3$)
As 
a matter of fact, 
for a container of dimensions $L\times L\times L$, a particle
of mass $m$ and speed $v_x$ moving along the $x$ direction hits one of 
the walls $v_x/2L$ times per unit time, and in each collision 
transfers a momentum $2 m v_x$. Thus, the momentum transfer per unit time
to the wall is $m v_x^2/L$. If we have $N$ particles moving along the $x$ direction
and with different speeds, the average momentum transfer per unit time is 
$\langle \Delta p / \Delta t\rangle = 
N m \langle v_x^2\rangle/L$, where $\langle v_x^2\rangle$ is the average value
of $v_x^2$. Further assuming that there is no privileged direction of motion,
$\langle v^2\rangle = \langle v_x^2 + v_y^2 + v_z^2\rangle = 3\langle v_x^2\rangle$.
Finally, dividing by the area $L^2$ to obtain the pressure,
\begin{equation}
P = \frac{1}{L^2}\left \langle \frac{\Delta p}{\Delta t}\right\rangle = 
\frac{1}{3} \frac{N}{V}m \langle v^2\rangle\ ,
\end{equation}
which is Equation~(\ref{pressure}) identifying the internal energy
with the kinetic energy of the gas. Take note that the potential energy of the
constituents of the gas is negligible in regard to their kinetic energy. Furthermore,
all the main results derived below do not depend on this approximation.

It is worth noting that, \textit{despite the absence of friction}, \textit{the piston does reach a final 
equilibrium position} and does not
remain oscillating indefinitely. Or, more precisely, will at first oscillate while slowly evolving
to its equilibrium position and, once this position is reached, the piston stays 
\textit{jiggling} around it. 
What happens is that the pressure exerted by the gas on the piston is 
a ``dynamic pressure''. \cite{Abreu1996,Abreu2002b,Gislason2010,AG2011}  For the
same volume, the dynamic pressure is higher in a compression and lower in
an expansion.\cite{Abreu1996,Abreu2002b,Gislason2010} The piston moves under the
action of the weight and of the dynamic force. Because of the
unbalance between both forces, the latter has a ``dissipative character,''
leading the piston to the final equilibrium position.\cite{Abreu2002b}
The situation is somewhat
similar to the case of a bullet entering at high speed in a region where there is a gas
at room temperature.
Even if all collisions between the gas particles and the bullet are perfectly elastic, the net effect is one 
of slowing down the bullet, until
it finally stops. We will return to this
point in Section~\ref{discussion}, where a one-dimensional version of our base model, 
interesting as a model of thought and adequate for numeric simulations,
is also suggested.

The final equilibrium position can be calculated from the principle of conservation 
of energy and the equality of the gas and piston pressures. The former corresponds
to state that the decrease in the gravitational potential energy must correspond
to an increase in the internal energy,
\begin{equation}\label{U2-U1}
-M g \left(h_2 -h_1\right) = U_2-U_1\ ,
\end{equation}
where the indexes $1$ and $2$ refer to the initial and final states, respectively, and $h$
is the height of the piston, $h=V/A$;
the latter is the statement
\begin{equation}
\frac{M g}{A} = \alpha \frac{U_2}{V_2}\ ,
\end{equation}
where Equation~(\ref{pressure}) was used. We have as well
\begin{equation}
U_1 = \frac{P_1V_1}{\alpha}\ ,
\end{equation}
so that
\begin{equation}\label{h2}
h_2 = h_1\frac{\alpha}{1+\alpha}\left(1+\frac{P_1 A}{M g \alpha}\right)\ .
\end{equation}
If $P_1< P_e$, as we have assumed, $h_2< h_1$, as it should be. 
Nevertheless, Equation~(\ref{h2})
is valid for any relation between the initial gas pressure $P_1$ and $P_e$.

Let us now assume we remove a certain mass from the piston (for instance,
we can imagine that
the piston has several masses on the top of it, and we can thus remove to the side 
one of them). The new total mass of the piston is $M^\prime<M$ and the piston
will reach a new equilibrium position, which we identify with the subscript 3. Let us 
look at the case where $M^\prime$ is
such that the new equilibrium position, $h_3$, is the same as the initial position $h_1$.
The energy conservation between positions 2 and 3 reads, similarly to
Equation~(\ref{U2-U1}),
\begin{equation}
-M^\prime g\left( h_1-h_2\right) = U_3-U_2\ .
\end{equation}
Therefore, combining this expression with (\ref{U2-U1}),
we arrive at the following very interesting result 
\begin{equation}\label{U3-U1}
U_3-U_1 = \left(M-M^\prime\right) g \left (h_1-h_2\right) > 0\ .
\end{equation}
Thereby, when the piston goes back to
its initial volume, the internal energy of the gas is higher than the initial one.
It is straightforward
to show that this conclusion remains valid if initially we would have $P_1>P_e$ 
and hence
$h_2>h_1$ and $M^\prime>M$.

Equation~(\ref{U3-U1}) means that the internal energy cannot be solely a function
of the volume, in contrast to the typical situations of mechanics
(where the potential energy 
is only a function of position). If we assume that one further variable suffices to
completely determine the internal energy, then we can write
\begin{equation}\label{U(S,V)}
U=U(V,S)\ .
\end{equation}
The new variable, $S$, is called \textit{entropy}.

\subsection{Properties of entropy and the second law}\label{propentropy}

Suppose now that the piston has initially a mass $M_0$ such that $ A P_1/g=M_0$,
which means that if the piston is released it stays at its equilibrium position. We can
now add to the piston a mass $m=(M-M_0)/2$ and wait for the piston to reach its equilibrium
position. Finally, we add yet another mass $m$ to the piston, so that its total mass is $M$, 
as in the compression described in Section~\ref{needentropy}.
Using the same reasoning
as above, it is not difficult to show that the final equilibrium position corresponds to
a position $h_2^\prime < h_2$. This, again, shows that the internal energy cannot be
uniquely a function of the volume. Moreover, the lower height reached by the piston
reflects the smaller unbalance between the dynamic force exerted by the gas and the
weight than in the previous case.

\begin{figure}
\begin{center}
\scalebox{0.80}{\includegraphics{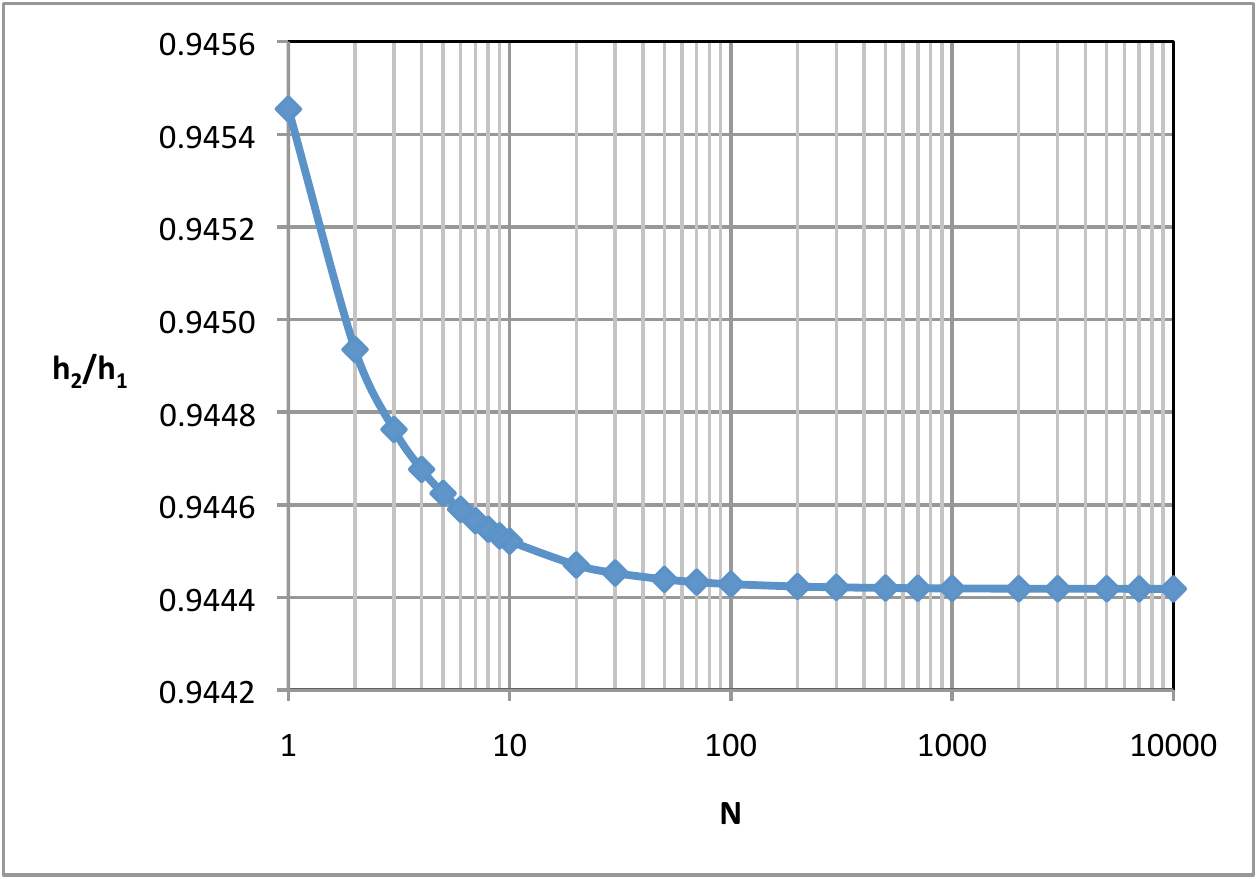}}
\caption{\label{sand}Final equilibrium height of the piston after the compression,
as a function of the number $N$ of masses used to increase the total mass from $M_0=10$ kg
to $M=11$ kg (see text).}
\end{center}
\end{figure}

We can further imagine a transformation where the mass of the piston is increased
in $N$ equal steps from $M_0$ to $M$, for instance by slowly adding sand grains, 
one by one, 
on the top of the piston. At each step the dynamic force exerted by the gas is nearly
the same 
as the weight. Therefore, the dissipative character of the dynamic force practically
does not manifest itself. The piston reaches the lowest possible final position corresponding
to a total mass $M$. Figure~\ref{sand} illustrates
this effect, by showing the final height of the piston as a function of the number $N$
of masses $m=(M-M_0)/N$ added on the top of  the piston. The calculations were carried
out for $M_0=10$ kg, $M=11$ kg and $\alpha=2/3$. The final height is depicted
as a fraction of the initial height.
The reader is invited to verify that the final result is given by 
$h_2/h_1=\prod_{k=1}^N\left[\frac{\alpha}{1+\alpha}+\frac{1}{1+\alpha}\frac{M_0+(k-1)m}{M_0+km}\right]$
and that the limiting result is $\left(M_0/M\right)^{1/(1+\alpha)}$. For the case depicted in
Figure~\ref{sand} we have $(10/11)^{3/5}\simeq0.944418$.

\begin{figure}
\begin{center}
\scalebox{0.80}{\includegraphics{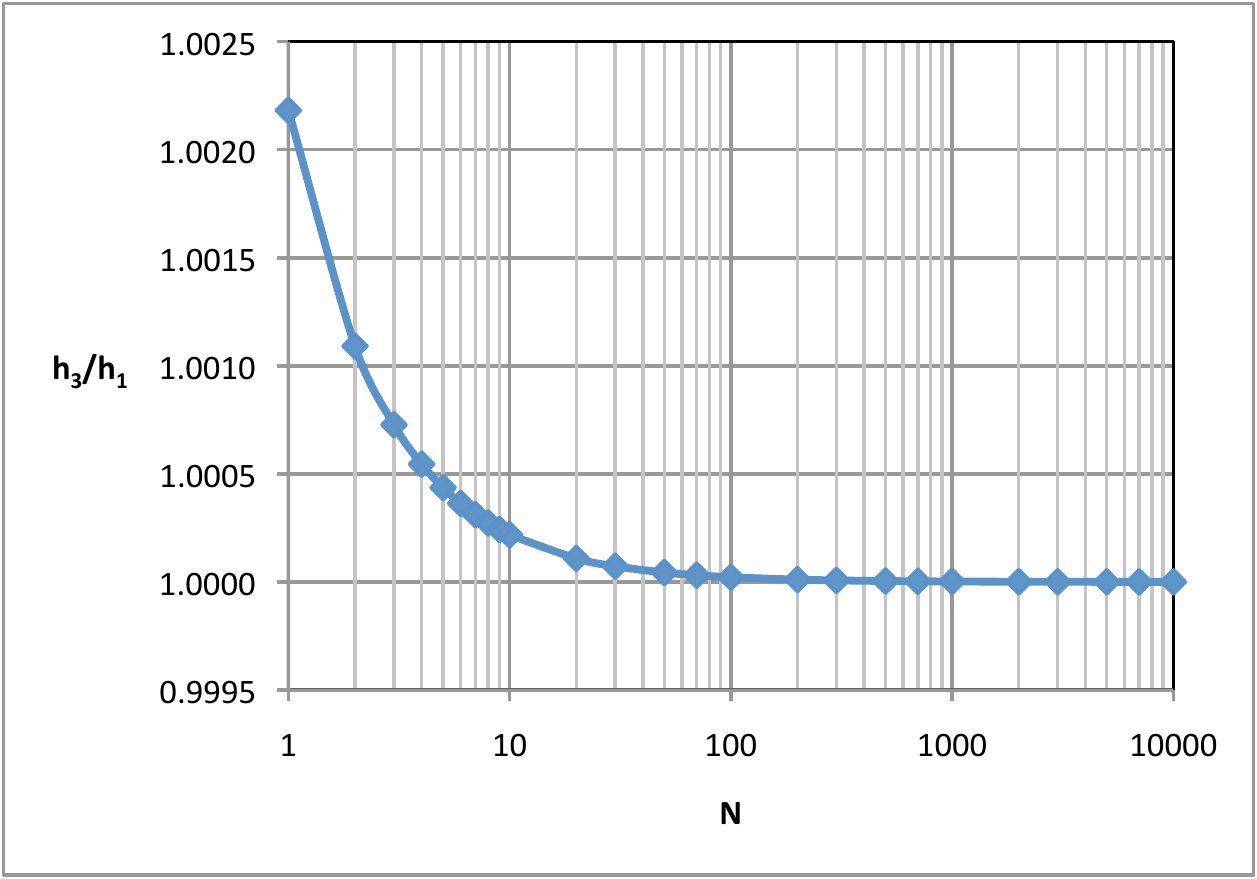}}
\caption{\label{reversible}Final equilibrium height of the piston after compression and
expansion, as a function of the number $N$ of masses used both to increase the total mass from $M_0=10$ kg
to $M=11$ kg and then to decrease it back to $M_0$ (see text).}
\end{center}
\end{figure}

Interestingly enough, if the sand grains are now slowly removed,
one by one, we have again almost an equilibrium between the dynamic force and the weight
at all times.
Everything nearly returns to the initial configuration and the gas will thus have very
approximately the same internal
energy as in the beginning. Figure~\ref{reversible} shows the final height $h_3^\prime$
as a function of masses used both during the compression and the expansion, normalised
to the initial height $h_1$.

Lastly, note that if we make the compression with a large number of steps, but the
expansion quickly in few steps, we still find the need to add some additional mass
to the piston to bring it to height $h_1$ and $U_3>U_1$, as in Equation~(\ref{U3-U1}).
The same happens if we do the expansion slowly with a large number of steps, but the
compression with few steps.

Some properties of entropy can now be easily derived. Referring to this simple case treated
with the base model, we have
\begin{equation}\label{S3S1}
U(V,S_3) \geq U(V,S_1)\ .
\end{equation}
If we arbitrate that $S_3\geq S_1$, then \textit{the entropy can only increase or
remain constant}. Moreover,
\begin{equation}\label{USV}
\left( \frac{\partial U}{\partial S}\right)_V >0\ .
\end{equation}
Clearly, the situation with no entropy change corresponds to a
\textit{reversible} transformation, \textit{i.e.}, to a transformation where the initial
state of the system can be recovered.
In this case the internal
energy can be calculated only from the volume, as in the typical cases of mechanics.
Thereby, mechanics corresponds to \textit{isentropic} thermodynamics, that is, to a set
of transformations where the entropy does not change.

It is worth underlining two issues evidenced by the previous discussion and
Figures \ref{sand} and \ref{reversible}. The first one is that
for deformable bodies (bodies with internal structure), the isentropic transformations can
only be performed by \textit{always} keeping the system nearly in equilibrium, when
the dynamic force is always approximately balanced by the ``static'' weight force. 
Failing to do so, \textit{either in the compression or the expansion}, or both, leads to an increase of the entropy and to the impossibility of recovering the initial state of the system. 
The latter statement exposes the second matter:
it is not necessary to look at the complete sequence compression+expansion to
speak about a reversible transformation,
as its reversible character, corresponding to constant entropy during the process, 
can be ascribed separately to 
the compression and to the expansion. 
In short, if 
a transformation leads from one state to a different one
with the same entropy, then the transformation can be reversed, and vice-versa.
The process just described of slowly adding or subtracting sand grains approaches such reversible processes.

We have thus obtained the second
law of thermodynamics, while keeping its interpretation and understanding at a very fundamental level.
Evidently, in many situations studied in mechanics a variation of entropy does take place, although usually it is not taken into account.
An example would be the treatment of the compression depicted in Figure~\ref{deform}a)
considering the energy only as a function of the deformation. A similar case is the deformation of a spring and the application of Hooke's law.
We will further discuss this point in Section~\ref{discussion}.

\subsection{Introduction of the formalism}

As shown in the previous Sections, the internal energy depends (at least) on two variables, volume
and entropy, as given by Equation~(\ref{U(S,V)}). Therefore, we can write
\begin{equation}\label{dU}
dU = \left(\frac{\partial U}{\partial V}\right)_S dV+ \left(\frac{\partial U}{\partial S}\right)_V dS\ .
\end{equation}

The idea of the increase in entropy as a result of the lack of balance between the
dynamic and the static force, advanced and discussed in the previous sections, can be
confirmed as follows. The work of the resultant of the forces on the piston is equal to
the variation of its kinetic energy. In differential form,
\begin{equation}\label{dEk}
dE_k = -\frac{Mg}{A} dV + \tilde{P} dV\ ,
\end{equation}
where $\tilde{P}$ is the dynamic pressure exerted by the gas on the piston and $dE_k$
is the variation of the kinetic energy of the piston. On the other hand, the variation of
the potential energy of the piston is
\begin{equation}\label{dEPot}
dE_{pot} = \frac{Mg}{A} dV\ .
\end{equation}
Consequently,
\begin{equation}\label{dEmec}
dE_k + dE_{pot} = \tilde{P} dV\ .
\end{equation}
In addition, the conservation of energy reads
\begin{equation}\label{EnergyCons}
dE_k + dE_{pot} + dU = 0\ .
\end{equation}
As a result, 
\begin{equation}\label{P'dV}
dU = - \tilde{P} dV\ ,
\end{equation}
so that \textit{the variation of the
internal energy of the gas is equal to the work done 
by the dynamic pressure}.\cite{Abreu2002b,Gislason2010, AG2011}

Among the conclusions from Section~\ref{propentropy} we have seen that a reversible transformation corresponds to
a constant value of the entropy, so that $dS=0$. What is more, the dynamic pressure
is the same as the static pressure in this case. Hence, as $dU=-P_e dV$ and
 $P=P_e$,
 with $P_e$ given by (\ref{Pe}), from (\ref{dU}) we have
\begin{equation}
dU = -P_edV=-PdV = \left(\frac{\partial U}{\partial V}\right)_S dV\ .
\end{equation}
Thus, the quantity $P$ can be defined from
\begin{equation}\label{P}
P = - \left(\frac{\partial U}{\partial V}\right)_S\ ,
\end{equation}
which corresponds to the gas pressure for an equilibrium point.

Similarly, we can define the quantity $T$ from
\begin{equation}\label{T}
T =  \left(\frac{\partial U}{\partial S}\right)_V >0\ ,
\end{equation}
where the inequality is simply Equation~(\ref{USV}). It is not difficult to later identify $T$ with the 
ideal gas temperature. This determination has been made by other 
authors.\cite{Callen1985,DeHeer1986}
Notice, however, that, for an irreversible transformation, $P$ and $T$ are
\textit{defined} by Equations~(\ref{P}) and (\ref{T}), respectively.
In a dynamic situation, when the gas
has a certain volume and a certain internal energy, $P$ and $T$ are the 
pressure and temperature \textit{it would have if it were
in equilibrium and would have the same volume and internal energy}.
This is the general meaning of $P$ and $T$, and no other. Furthermore,
in a dynamic situation $P$ is \textit{not} the
pressure exerted by the gas.
Naming $P$ and $T$ 
``pressure'' and ``temperature'' and thinking in physical terms in these
quantities, with these designations, as defined in equilibrium, is a common source 
of mistakes and misunderstandings. 

Finally, we can write
\begin{equation}\label{dU2}
dU = - \tilde{P} dV = -P dV + T dS
\end{equation}
and
\begin{equation}\label{PP'}
-(\tilde{P}-P)dV = TdS\ .
\end{equation}
This last equation establishes that \textit{the variation of entropy is a consequence
of the difference between the dynamic and the static pressures}. This difference results
in the ``dissipative character'' of the force, \textit{even if there is no friction}, as pointed out
and discussed in section
\ref{needentropy}.
This expression allows an additional verification that $S$ always increases. As a matter of
fact, if $dV>0$ it must be true that $\tilde{P}<P$, so that $dS>0$. The same conclusion is
obtained if $dV<0$, as then $\tilde{P}>P$.

To finish this section, let us go back to Equation~(\ref{P'dV}), to note that
\begin{equation}\label{DeltaU}
\Delta U = -\int \tilde{P} dV\ ,
\end{equation}
which, denoting by $W$ the work done by the dynamic force,
\begin{equation}\label{W}
W = -\int \tilde{P} dV\ ,
\end{equation}
takes the expected form
\begin{equation}
\Delta U = W\ . 
\end{equation}
Moreover, it is immediate to verify that between two points where the piston is at rest the work done by the dynamic force $W$ is equal to the work
of the weight. Indeed, substituting (\ref{dEPot}) in (\ref{EnergyCons}), integrating and noting that in this case $\Delta E_k=0$,
\begin{equation}
\Delta U = -\int \frac{Mg}{A}dV = -\int P_e dV = -\frac{Mg}{A}\Delta V = -\Delta E_p = W\ .
\end{equation}
This consistency check does not constitute a surprise, as all our analysis of the base model started precisely from this condition.
The interesting fact is that, despite $P_e\neq \tilde{P}$ along the transformation, between two points where the piston is at rest we nevertheless have
\begin{equation}
\int P_e dV= \int \tilde{P} dV\ .
\end{equation}

If the transformation is reversible, $dS=0$ along the transformation and, if the gas
returns to the initial volume, using Equation~(\ref{dU2}),
\begin{equation}
\Delta U = - \oint \tilde{P} dV = - \oint P dV = 0\ .
\end{equation} 
In general, for an irreversible transformation returning to the same volume,
\begin{equation}
\Delta U = - \oint \tilde{P} dV >0\ .
\end{equation}
Thus, the work of the dynamic force is transformed in internal energy, making it increase
if the system returns to the initial volume, in accordance with the conclusions
presented in
Section~\ref{needentropy}. 

In a generalisation of the concept of heat,\cite{Brush1986} we can say that
the work has been transformed into \textit{heat}. Heat, however, is is a subtle concept, still 
often misinterpreted (\textit{cf.} Section~\ref{adiabaticpiston}) and it would be better
to rather say that ``work has been used to heat the gas.'' Nonetheless,
with the intuition acquired on the notion of variation of entropy associated with the base model,
linking Newton's second law with the second law of thermodynamics,\cite{Abreu2002b} 
we are now ready to infer the properties of more complex systems.

\section{Heat: generalizations of the base model}\label{Generalizations}

The natural generalization to a more elaborated configuration is the study of 
a system comprised of 
two subsystems with a common boundary, from which they can exchange energy.
We will consider various
configurations. The first one corresponds to one subsystem of fixed volume with the
second subsystem with a movable piston on top, leading to the formulation of the first law and the study of the heat reservoir. 
The second arrangement
is the case of two subsystems side by side coupled by a moving piston, which allows 
the analysis of the celebrated ``adiabatic piston'' problem.
\cite{Feyetal1979,Kesetal2000,Abreu2002,Manetal2006,Gislason2010,AG2011}
A third geometry is the same as the previous one, but with the two subsystems on the vertical. 
Finally, the last setup
involves two-subsystems side by side and with a piston on the top of each of them, in order to
study the constant pressure calorimeter.\cite{GC2005,GC2007}

\subsection{Heat and the first law}\label{1stlaw}

\begin{figure}
\begin{center}
\scalebox{0.15}{\includegraphics{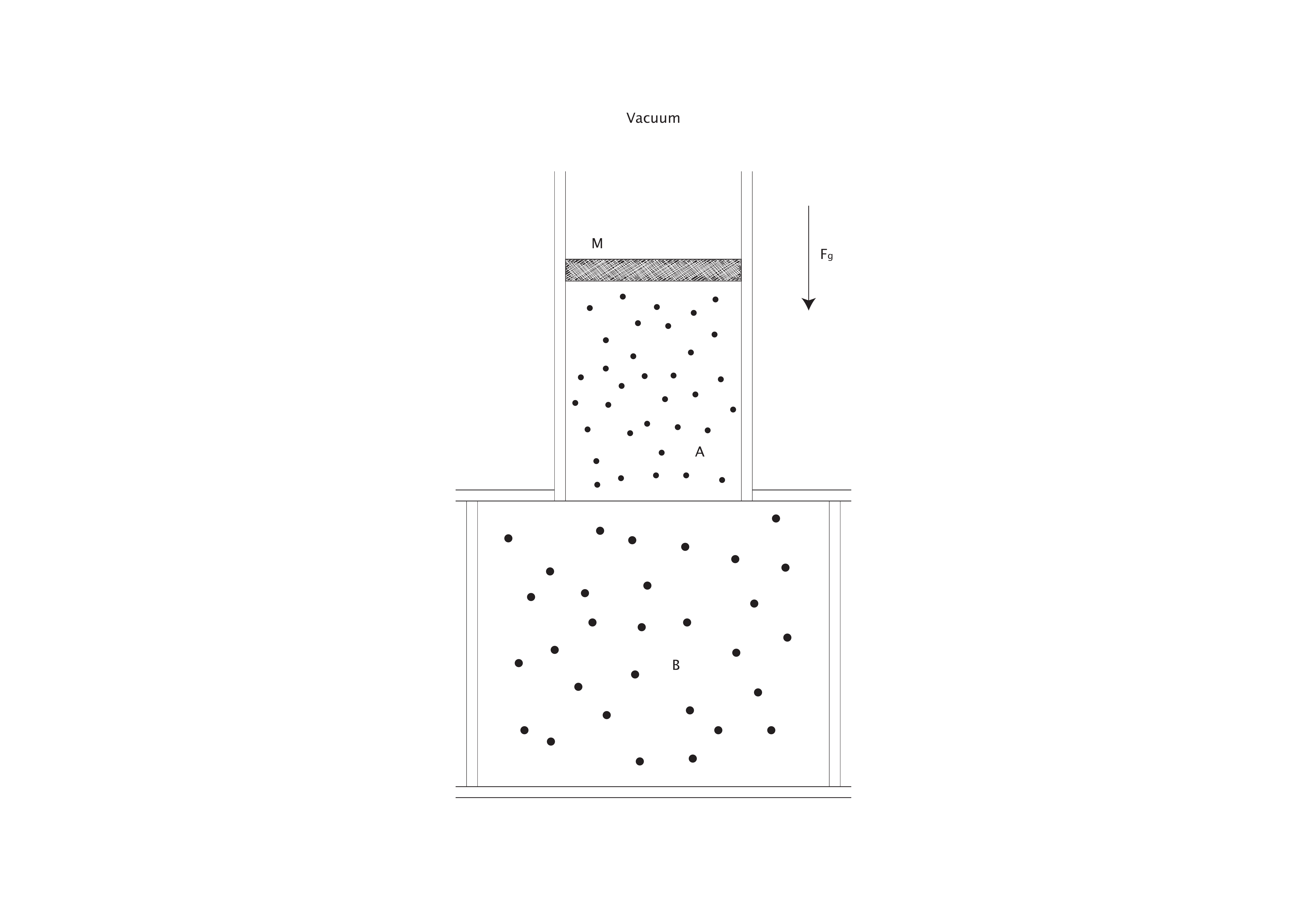}}
\caption{\label{Gener1}Two subsystems, $A$ and $B$,
which can exchange energy through a common border. A frictionless piston of mass
$M$ can move and modify the volume of subsystem $A$.}
\end{center}
\end{figure}

The first geometry we want to investigate corresponds to a subsystem $A$ similar to the
base model, namely a classical ideal gas in a container with a moving piston.
However, subsystem $A$ is in now contact with another subsystem $B$, of fixed volume,
as represented in Figure~\ref{Gener1}.
The \textit{complete} system (formed by gases $A$ and $B$ and the piston) is isolated, but subsystems  $A$ and $B$ can exchange energy
with the piston and therefore the system ($A+B$) is not isolated. 
Furthermore, subsystems $A$ and $B$ can also change energy 
through the common border. 
The conclusions from our base model apply to the new system ($A+B$).

As the complete system is surrounded by vacuum, we have
\begin{equation}\label{UAUB}
U = U_A+U_B\ ,
\end{equation}
where $U$ is the total internal energy, and $U_A$ and $U_B$ are the internal
energies of subsystems $A$ and $B$, respectively.

We can now write [\textit{cf.} Equations~(\ref{DeltaU})--(\ref{W})]
\begin{equation}
W = -\int \tilde{P} dV = \Delta U = \Delta U_A + \Delta U_B\ .
\end{equation}
As before, between two points where the piston is at rest the work of the dynamic force is equal to
the work of the weight.
The last expression can be written as
\begin{equation}
\Delta U_A = W - \Delta U_B\ ,
\end{equation}
or, defining
\begin{equation}\label{DUB}
Q = -\Delta U_B\ ,
\end{equation}
\begin{equation}\label{1stP}
\Delta U_A = W + Q\ .
\end{equation}
Evidently, this is the usual formulation of the first law of thermodynamics 
(the symbol $W$ is often defined with the symmetric sign, \textit{i.e.}, as the work
done by the system, so that $\Delta U_A =  Q - W$). The
quantity $Q$ is denoted by \textit{heat} or heat exchanged with subsystem $B$.
Its symmetric simply 
corresponds to the \textit{variation of the internal energy of subsystem $B$} 
or, similarly,  the energy transferred from $A$ to $B$.
The first law is thus a particular form of writing the principle of conservation of energy.

It is important to emphasize that the introduction of heat in this simple but clear way 
gives a valuable contribution to help avoiding some of the most common traps 
related to this concept, as further discussed in Section~\ref{adiabaticpiston}.

\subsection{Temperature and the heat reservoir}\label{heatreservoir}

One interesting result that can be easily obtained is the equality of temperatures
of subsystems $A$ and $B$ described in Section~\ref{1stlaw}
along reversible transformations. In fact, from (\ref{UAUB})
\begin{equation}
dU=dU_A+dU_B\ .
\end{equation}
Here, noting that $V_B=$\textit{const.} and that for a reversible transformation $dS=0$\ ,
\begin{equation}
dU = \left(\frac{\partial U}{\partial S}\right)_V dS - P dV = - P_A dV_A\ .
\end{equation}
On the other hand,
\begin{equation}
dU_A = T_A dS_A - P_A dV_A\ ,
\end{equation}
and
\begin{equation}\label{dUB}
dU_B = T_B dS_B\ ,
\end{equation}
where $T_A$ and $T_B$ are defined by Equation~(\ref{T}).
Combining these expressions, we obtain
\begin{equation}\label{dSAdSB}
T_A dS_A + T_B dS_B = 0\ .
\end{equation}
Finally, assuming the entropy to be an extensive quantity (an assumption that can be
motivated and discussed) and since the transformation is reversible,
\begin{equation}
dS= dS_A+dS_B= 0\ ,
\end{equation}
from where, using (\ref{dSAdSB}),
\begin{equation}\label{TATB}
T_A = T_B\ ,
\end{equation}
which is the result we wanted to achieve. It shows that temperature is a property
characterizing the state of equilibrium of the two subsystems.

The limiting case in which the subsystem $B$ is infinite is important, as it 
coincides with the idea of a small subsystem in contact with a \textit{heat reservoir}.
In this case, subsystem $B$ is so big that an energy exchange with the smaller
subsystem does not change its temperature. 
This intuitive notion can be verified as follows. If the energy density is uniform, in each unit volume $i$ of the heat reservoir we have
the same volume $V_i$ and the same internal energy $U_i$. Hence, since $U_i=U_i(S_i,V_i)$, 
all unit volumes have as well the same $S_i$.
As the system is infinite, any finite transfer of energy
will not change its energy per unit volume.  Therefore, $V_i$, $U_i$ and $S_i$ are not modified by a finite energy transfer to or from the heat reservoir.
All quantities being the same, the derivatives (\ref{T}) are also the same and the temperature of the heat reservoir always
remains unchanged.
Equation~(\ref{TATB}) tells us that
in this case a reversible transformation is an isothermal transformation at $T=T_B$. Moreover,
since $dS_A=-dS_B$, using (\ref{dUB}) and (\ref{DUB}),
\begin{equation}
dS_A = \frac{dQ}{T}\ .
\end{equation}

\subsection{The adiabatic piston}\label{adiabaticpiston}

\begin{figure}
\begin{center}
\scalebox{0.15}{\includegraphics{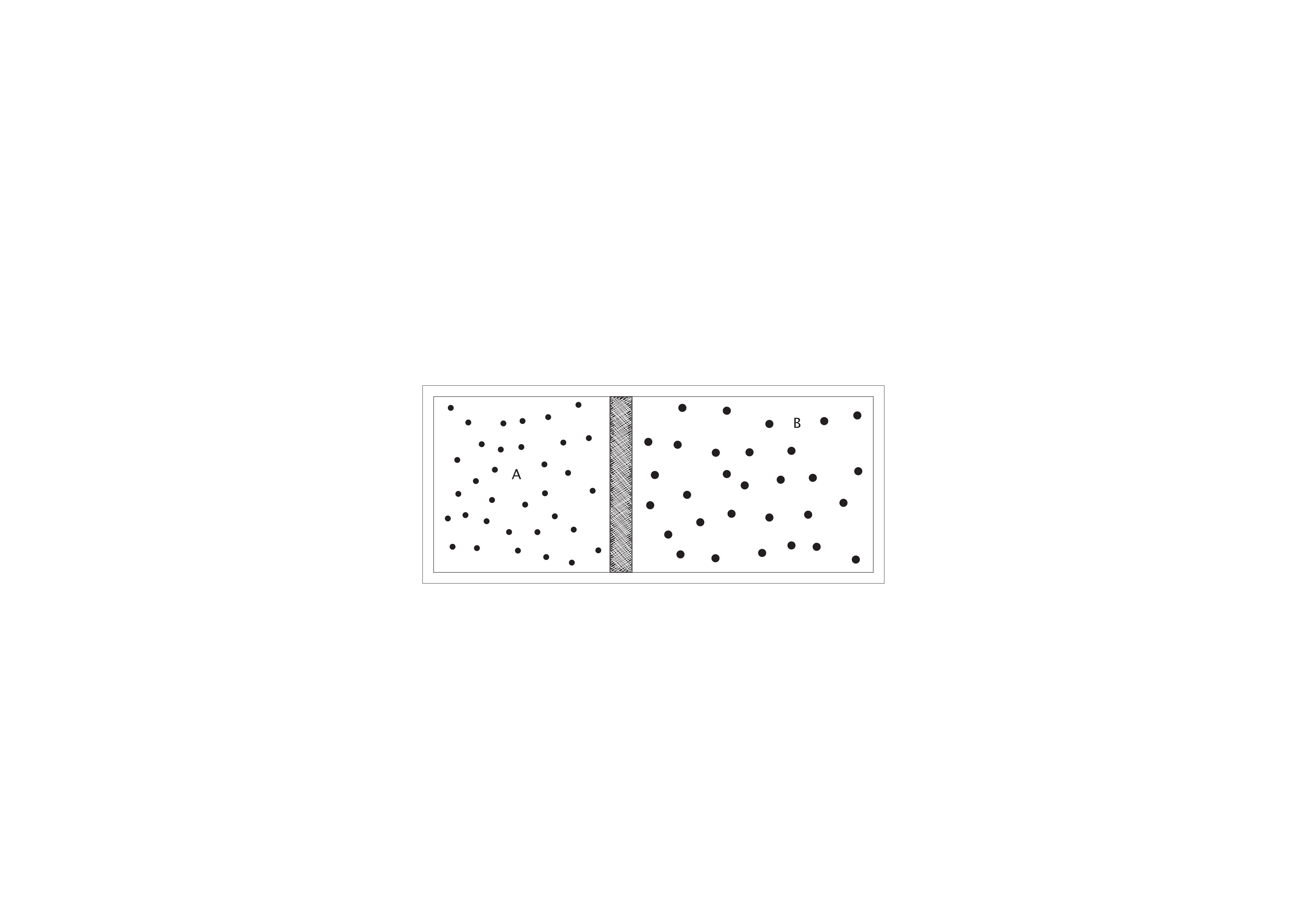}}
\caption{\label{AdiabaticPiston}Two subsystems, $A$ and $B$,
which \textit{can} exchange energy through a moving ``adiabatic'' and frictionless piston.}
\end{center}
\end{figure}

The so called ``adiabatic piston'' problem corresponds to a system involving two ideal gases contained in a horizontal cylinder and separated
by an \textit{insulating piston} that \textit{moves without friction}, as shown in Figure~\ref{AdiabaticPiston}. 
This system is more complex than the previous ones and was treated by many 
authors, using different approaches and techniques.\cite{Feyetal1979,Kesetal2000,Abreu2002,Manetal2006,AA2008,BK2010,Gislason2010,AG2011}
Worth noting are the qualitative kinetic description by Feynman,\cite{Feyetal1979} the molecular dynamics calculations by Mansour and 
co-workers\cite{Kesetal2000,Manetal2006} and the classical thermodynamics analysis by Gislason.\cite{Gislason2010}

Classical thermodynamics analyses are of major interest here. As pointed out in our previous work,\cite{AG2011} a careful use of thermodynamics must give the same final
result as molecular dynamics, because the latter is a microscopic interpretation of the former. However, too commonly this is not the case,
a fact that strikingly exemplifies the difficulties associated with the formalism of thermodynamics.
Gislason makes a 
very interesting and enlightening discussion of the problem,\cite{Gislason2010} focusing on the shorter 
time-scale, when the two gases evolve to a situation of equal pressures. Though, 
he does not address the
second phase, when the gases evolve to a situation of equal temperatures, discussed qualitatively by Feynman\cite{Feyetal1979} and 
formally derived in our previous work.\cite{AG2011} 
On the other hand, Anacleto and Anacleto,\cite{AA2008}
just to give one example, make a faulty investigation of the problem, claiming that the piston does not reach a final state of equilibrium, instead
keeping oscillating indefinitely. Furthermore, they allege that entropy remains constant, due to the absence of friction, which is not the case.

The main difficulty with this problem lays in a negligent use of language.
As a matter of fact, the word ``adiabatic'' is too swiftly associated with ``no heat exchange.'' Moreover, ``heat exchange'' is rarely defined with generality, but
we immediately are lead to impose the mathematical condition $dQ=0$ in the calculations. 

The correct solution within the framework of thermodynamics was presented previously,\cite{Abreu2002,AG2011} and the reader should
refer to those papers for the details on the formal use of the thermodynamic laws.
Quoting from our former work,\cite{AG2011} by an ``adiabatic piston''
it is meant a piston \textit{with zero heat conductivity}.
If the piston is held in place (for instance, if it is fixed to the box by screws), then there is no
``heat transfer''  from one subsystem to the other. Even though, if the piston is released, both systems 
exchange energy via collisions with the piston, as they
are coupled
through the conditions of 
constant total volume and total energy, where the kinetic energy of the piston has to be taken into account. 
The evolution to a stage of mechanical equilibrium of equal pressures \textit{has nothing to do with friction}: it is simply a result of 
the dissipative character of the dynamic pressure, discussed in 
our base model. There is indeed an entropy increase, as also acknowledged by Gislason in his analysis of the first phase of the problem.\cite{Gislason2010}
Gislason in fact provides significant physical insight by identifying the
damping of the piston motion as a result of the dynamic pressure on it, ``because the pressure is
greater when the piston is moving towards the gas than when the piston is moving away from 
the gas.'' \cite{Gislason2010} 

After the equalization of pressures the coupling between both subsystems remains, only the kinetic energy of the piston becomes negligible.
Still, the collisions between the gas particles and the piston
will make the piston jiggle, \textit{allowing an exchange of energy} between both gases.\cite{Feyetal1979,AG2011}
These energy exchanges will always take place, despite the piston being non thermal conductor and the absence friction, as they are
simply a result of the momentum transfer in the collisions (\textit{cf.} the discussion by Feynman\cite{Feyetal1979}).
And, as pointed out in Section~\ref{1stlaw}, these energy exchanges can be formally treated as heat exchanges.
Therefore, in this second phase the system evolves to a situation
of equal temperatures, with $\Delta U_A + \Delta U_B = 0$. In this case, if we write the first law for gas $A$ we 
have $Q=-\Delta U_B\neq0$ [Equation~(\ref{dUB})],
and \textit{the condition $dQ=0$ cannot be imposed}.\cite{AG2011} In fact, we have instead $dQ_A=-dQ_B$
[equation (16)
in Ref.\cite{AG2011}]. Notice that the different quantities somewhat lose their energetic interpretation, merely being a result of the mathematical formalism.

From the discussion above, it is clear that during both phases of evolution there \textit{has to be} an ``heat exchange,'' according to the
formalism of thermodynamics, no matter wether the piston was 
defined as ``adiabatic,'' which might seem a bit shocking at first.
The problematic use of language is easily avoided if
we leave behind a formulation of the first law which to some extent still dates from the time of caloric, and instead keep in mind its introduction
as suggested in Section~\ref{1stlaw} and from Equation (\ref{DUB}).
Then there is no doubt that the ``adiabatic piston'' system allows the
energy exchange between both subsystems. And it is by no means shocking to assert that the internal energy of each subsystem changes due to the collisions, 
even for a piston with zero heat conductivity and moving without friction. 

\subsection{Further generalizations}

From a general introduction of the first and second laws of thermodynamics and an early alert on the dangers of a blind use of the mathematical formalism,
as outlined in the previous sections, it is possible to proceed to more complicated and richer systems. This paper would become too lengthy if we
would discuss them in detail here. Nevertheless, due to their importance we find justified to delineate few of them already. 

\subsubsection{The adiabatic piston in a gravitational field}

A natural generalization is to consider the adiabatic piston from the previous section, but now in a vertical configuration and under the effect of gravity. 
In this case the work done by the dynamic force has two terms, one
for the each gas. This case follows very closely the adiabatic piston discussed in our previous work.\cite{AG2011}

The conservation of energy reads
\begin{equation}\label{dEpdEk}
dU_A+dU_B+dE_{pot}+dE_k=0\ ,
\end{equation}
where $dU_A$ and $dU_B$ are the internal energies of subsystems $A$ (bottom) and $B$ (top), respectively, whereas $dE_{pot}$
and $dE_k$ are the piston gravitational potential energy and its kinetic energy, respectively. 
If the piston has mass $M$ and area $S$, and noting that $dE_{pot}=(Mg/A)dV_A$,
the reader is invited to adapt our former calculations\cite{AG2011} and verify that the equilibrium condition corresponds to equality of forces on 
the piston \textit{and} equality
of temperatures, $P_A=P_B+Mg/A$ and $T_A=T_B$.

Furthermore, since $dV_A+dV_B=0$,
the work of the resultant of the forces on the piston is [\textit{cf.} (\ref{dEk})]
\begin{equation}\label{dWAB}
(\tilde{P}_A-\tilde{P}_B) dV_A - \frac{Mg}{A} dV_A = dE_k\ ,
\end{equation}
Therefore, from (\ref{dEpdEk}) and (\ref{dWAB}) we have
\begin{equation}\label{dEkdEp}
(\tilde{P}_A-\tilde{P}_B)dV_A = dE_k+dE_{pot} = -dU_A-dU_B\ .
\end{equation}
Finally, 
\begin{equation}\label{dUi}
\sum_i dU_i = \sum_i\left(-\tilde{P}_i dV_i \right)= \sum_i\left( -P_i dV_i + T_i dS_i \right) \neq \sum_i\left(-P_i dV_i\right)
\end{equation}
with $i=\{A,B\}$ and
\begin{equation}
\sum_i T_i dS_i =-\sum_i(\tilde{P}_i-P_i) dV_i\ .
\end{equation}
Exactly like the case of the ``adiabatic piston,''
the direct use of the first law for one of the gases and the assignment of physical meaning to the quantity $Q$ is not straightforward,
as neither the conditions $dQ=0$ and $dU_i=-P_i dV_i$ nor even  $dU_i=-\tilde{P}_i dV_i$ can be imposed.\cite{AG2011,AG2010}

\subsubsection{The constant pressure calorimeter}

\begin{figure}
\begin{center}
\scalebox{0.15}{\includegraphics{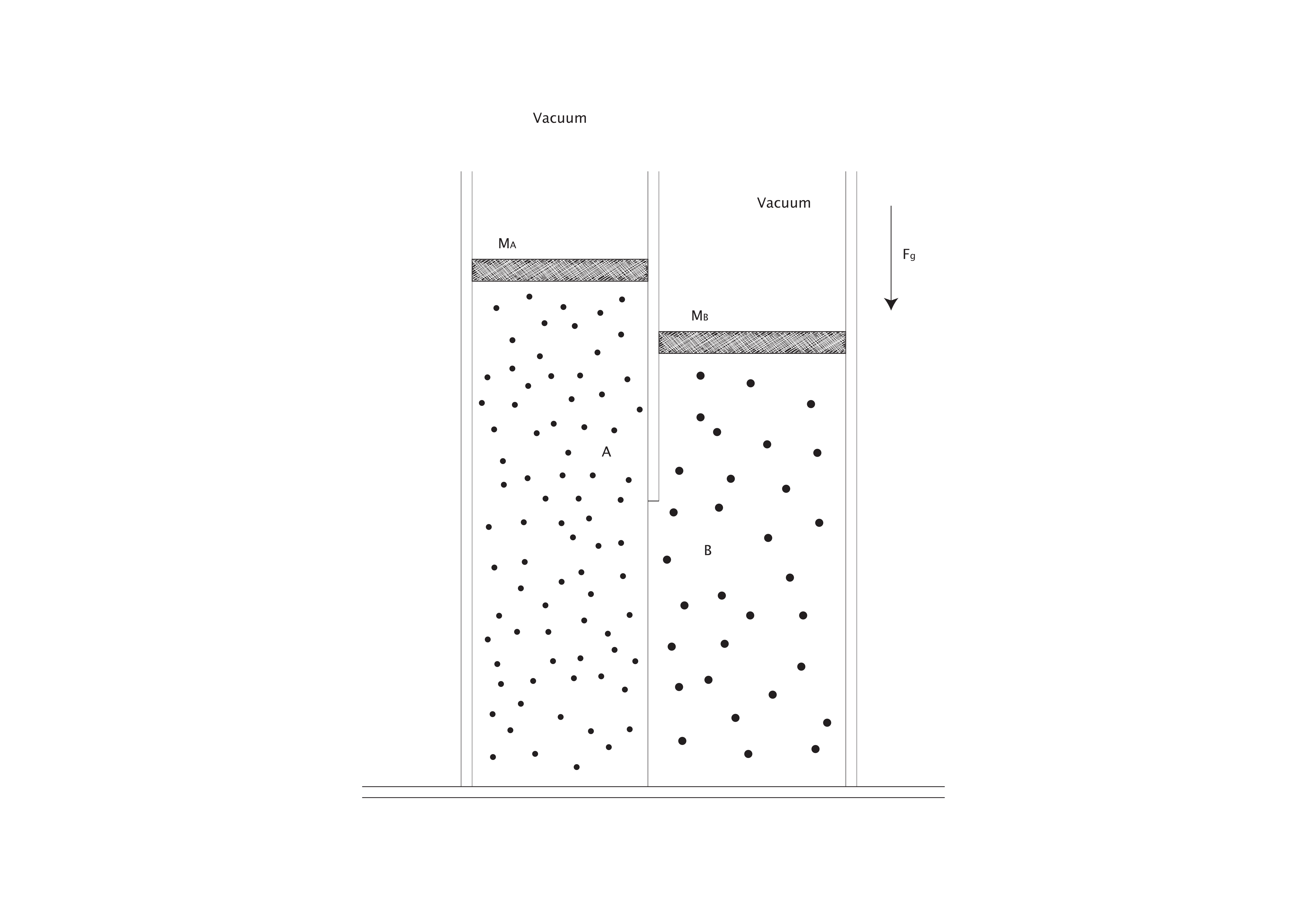}}
\caption{\label{Gener2}The constant pressure calorimeter can be schematically represented by two subsystems, $A$ (gas) and $B$ (calorimeter),
which can exchange energy through a common border. A frictionless piston of mass $M_B$
keeps subsystem $B$ at constant external pressure (instead of constant volume, as in Figure \ref{Gener1}).}
\end{center}
\end{figure}

One important configuration in practical applications is the constant pressure calorimeter.\cite{GC2005,GC2007}
The system can be idealized by adding a piston to subsystem $B$ from Figure~\ref{Gener1} to keep it at constant pressure.
The new configuration is depicted in Figure~\ref{Gener2}.

Let us denote the pressure exerted by the piston on subsystem $B$ by $P_0$. In other words, $P_0=M_Bg/A_B$,
where $M_B$ is the mass of the piston on subsystem $B$ and $A_B$ its area. Likewise, let us define
$P_e=M_Ag/A_A$. 

Consider first a reversible transformation. In this case, the pressure $P_B$ is always equilibrated  at $P_0$ and
there is no difference between $P_B$ and $\tilde{P}_B$. Therefore, defining the enthalpy $H$ as
\begin{equation}
H = U + PV
\end{equation}
and the specific heat at constant pressure $C_P$ from
\begin{equation}
(dH)_P = C_P dT\ ,
\end{equation}
we have
\begin{equation}
dH_B =  d(U_B + P_B V_B) = dU_B + P_B dV_B = T_B dS_B \equiv C_{P,B} dT_B\ .
\end{equation}

To address the general case of an irreversible transformation (for instance if we would initially have $P_e>P_A$),
we note that the conservation of energy reads
\begin{equation}
dU_A+dU_B+dE_{pot,A}+dE_{k,A}+dE_{pot,B}+dE_{k,B} = 0\ ,
\end{equation}
where $dE_{pot,A}$ and $dE_{k,A}$ are the potential and kinetic energies of piston $A$, 
given by Equations~(\ref{dEk}) and (\ref{dEPot}), respectively, and the same for piston $B$.
Hence, we still have
\begin{equation}
dU_A+dU_B = -\tilde{P}_A dV_A - \tilde{P}_B dV_B\ .
\end{equation}
Between two points where both pistons are at rest ($\Delta E_{k,A}=\Delta E_{k,B}=0$) we have, successively, 
\begin{equation}
\Delta U_A + \Delta U_B = -P_e\Delta V_A - P_0\Delta V_B\ ,
\end{equation}
\begin{equation}
\Delta U_A = -P_e\Delta V_A - (P_0\Delta V_B + \Delta U_B) =  -P_e\Delta V_A - \Delta H_B\ ,
\end{equation}
where $\Delta H_B = \int C_{P,B} dT$. If the specific heat at constant pressure $C_{P,B}$ is constant, we can finally write
\begin{equation}
\Delta U_A = -P_e \Delta V_A - C_{P,B} \Delta T_B\ .
\end{equation}
Thus, if we want to apply the first law of thermodynamics (\ref{1stP}) to gas $A$, the second term on the r.h.s. corresponds
to an energy exchange with subsystem $B$ that we can identify with 
the heat exchanged with subsystem $B$ between two points of equilibrium.

%
%

\section{Discussion}\label{discussion}

We have presented a simple and clear model to introduce thermodynamics, which reveals and naturally solves some of the difficulties 
underlaying the concepts of work and heat
in the formulation of the laws of thermodynamics. 

The first step is 
the extension of the notions of kinetic and potential gravitational energies to
the one of \textit{internal energy}, inferred in Section~\ref{WorkEnergy}
from the analysis of extended
deformable bodies.
Subsequently, the base model presented in Section~\ref{base} shows that if we
assume that part of the universe has potential and kinetic energy which depend only on the position and velocity, as in the study of \textit{mechanics},
then the remaining part of the universe -- in this case the gas -- has an energy which depends on the position and entropy, $U=(S,V)$, the
subject of \textit{thermodynamics}. 

It is pointed out that the \textit{dynamic force} on the piston has a \textit{dissipative character}, even if we have only conservative forces
and there is no friction. The second law of thermodynamics is then readily obtained (sections \ref{needentropy} and \ref{propentropy}).
An interesting idealized situation of our base model corresponds to
a very simple one-dimensional picture, namely, 
a gas formed of $N$ punctual particles of mass $m$ moving only on the vertical direction under the action of gravity, and colliding elastically
with the piston of mass $M$. There is no friction and the particles do not interact directly with each other. 
Even this straightforward model is enough to understand the dissipative character of the dynamic force, the
approach to equilibrium and, thus, irreversibility. In the thought case where all particles are initially exactly at the same height and have exactly the same velocity,
the situation is the same as with a one-dimensional elastic collision between two point masses (one of mass $Nm$ and the other of mass $M$). Therefore,
the piston remains oscillating indefinitely.
The dissipative character of the dynamic force does not appear
and the entropy remains constant. The system ``has no imagination,'' the accessible volume in phase-space
remaining very limited. However, if the masses $m$
are not exactly ``on phase,'' if there is a small difference in their positions or speeds, the dissipative character emerges and there is
an entropy increase.\endnote{It can be noted that the notion of ``exactly the same height and exactly the same velocity'' does not make
sense in quantum mechanics. However, it is not necessary to invoke quantum mechanics for the point we are making here.}
The accessible volume in phase-space has now increased. 
The key factor leading the evolution to equilibrium is the \textit{interaction} between the different particles, even if it is kept to a minimum
and only takes place indirectly through the collisions they experience with the piston.
These ideas are in line with the pioneering works by Ludwig Boltzmann.
A somewhat poetic statement expressing this main result would be ``thermodynamics is mechanics
with imagination.'' 


We believe that computer simulations are a powerful tool to explore the qualitative behavior of a model, as underlined by 
Gould and Tobochnik.\cite{GT2010a,GT2010b}
A systematic exploration of simulations of the simplified one-dimensional model just described, 
including the analysis of the accessible phase-space and equipartition of energy, will be performed in a future publication.
A very interesting simulation of a similar system, where collisions between particles and motion in two dimensions are considered -- but not the 
gravitational force on the particles --
is available online from the
NetLogo Models Library.\cite{Wilensky1997,Wilensky1999}
In this case, as the interactions between particles are easier to occur, the evolution to equilibrium is much faster than in our 
simplified one-dimesional case and the effect is
easily obtained.

The approach to equilibrium raises the question of \textit{irreversibility}. Still using the base model, irreversibility and the increase in entropy are seen to be
a result of the dissipative character of the dynamic force discussed above.
The example of the compression/expansion with sand grains illustrates that
the higher the unbalance between the dynamic and static pressures, the higher the increase in entropy, as also shown by Gislason.\cite{Gislason2010}
The importance of the so-called ``quasi-static'' formulations is then easily understood. It is the work of the dynamic force that is equal to the variation of the internal energy of the
gas [Equation~(\ref{P'dV})]. In a quasi-static process the work of the dynamic force is a good approximation for the work of the static force
during part of the trajectory of the piston, the
variation of entropy being nearly zero [Equation~(\ref{PP'})].
For any real process it is not possible
to actually return to the initial conditions and it is thus
necessary to generalize the idea of potential energy from mechanics, which is a function only of the variable of configuration 
(or deformation), $V$, to the gas internal energy, function of one additional variable, $S$.
It is noted that mechanics corresponds to isentropic thermodynamics, \textit{i.e.}, to the situations where $U=U(V)$ (and so the internal energy
is a potential energy), either by the nature of the problem or
as an approximation.

The traditional development of thermodynamics defines the internal energy first as a function of $V$ and $T$. Noting that 
$\left(\partial U/\partial T\right)_V=C_V>0$ we could then be induced to think
that mechanics would correspond to isothermal thermodynamics. However, this is not the case. During the sand-grain transformation of the base model, 
where $P\simeq\tilde{P}$, we have $dT\neq0$. 
In particular, during the compression and the expansion we have, respectively, $dT>0$ and $dT<0$. The transformation is thus characterized by $dS=0$
and not by $dT=0$. That being so, mechanics indeed corresponds to isentropic thermodynamics and not to isothermal thermodynamics, reinforcing the 
importance of considering $S$ the conjugate variable for $V$.

The idea that when a system returns to the initial position, such as after the compression-expansion from our base model, it has an higher internal energy 
than at the beginning [Equation~(\ref{U3-U1})], is fairly counterintuitive. 
This comes from the fact that physical systems are often surrounded by a thermostat, which prevents the manifestation of
the thermodynamic phenomenon. A good example is the deformation of an elastic material or a spring, such as the
case of a spring hanging on the vertical holding a certain mass. 
The situation is very much alike to our base model and a similar analysis as in Section~\ref{needentropy} can be made. 
New masses can be added and removed to the spring and new equilibrium positions can be found. 
In this case, the excess energy is exchanged with the surroundings
 and apparently we return to the initial state. The entropy variation is in the surrounding environment and we are 
 conducted to think that everything happens as if the dynamic force could be approximated by the static force and internal dissipation would not exist, as 
the spring returns to the initial length and the force to its initial value. 
However, we know that the energy increase of the environment is equal to the work of the dynamic force, which is equal
to the changes of gravitational potential energy of the masses that are now at a lower level. 

The heat reservoir was analyzed in Section~\ref{heatreservoir}, as a limiting case of the case of energy exchanges between two subsystems, which can
be described as \textit{heat} exchanges. The first law is also derived in this context (Section~\ref{1stlaw}).
Furthermore, 
the notion that ``there is no such thing as perpetual motion with weight lifting machines''\cite{Feyetal1979,Leonardo} (Section~\ref{WorkEnergy})
can now be easily extended to account for the second law.
In fact, this statement reflects the conservation of energy when entropy is not involved, so that $U=U(V)$. In this case, $W=\Delta U$ and,
when the system returns to its initial position, $\Delta U=0$ and hence $W=0$. For a general case, with friction or even simply the
reorganization of the internal energy as a result of the action of the dynamic force, $U=U(S,V)$, with $\Delta S>0$. 
When the system returns to its initial position we have $\Delta U>0$, so that 
$W>0$,
leading to the conclusion that ``there is no perpetual motion at all.''\cite{Feyetal1979}
This inference is valid both when subsystem $B$ 
is finite (\textit{cf.} Figure~\ref{Gener1}) and in the limiting case of an heat reservoir, in what may be seen as a generalization
of the Kelvin-Plank formulation of the second law.


The analysis of the adiabatic piston problem --  Section~\ref{adiabaticpiston} -- has to be done with care. It is no longer possible to separate 
the energy-momentum exchanges of
the particles from the two subsystems with the piston into quantities ``work'' and ``heat'' with clear energetic meaning. The correct and complete
solution of this problem may contribute to illustrate the difficulty in assigning a physical meaning to these two quantities, as they appear in the laws
of thermodynamics.\cite{Abreu2002,AG2011} 
The ``jiggling piston'' further provides a perfect bridge between thermodynamics and the microscopic structure of 
matter, Feynman's \textit{atomic hypothesis} mentioned in
the introduction to this paper,\cite{Feyetal1979} and the importance of the keeping it in mind.
Although the final result of equal pressures and temperatures can be obtained without referring to heat and thermodynamics,\cite{Feyetal1979}
the complete analysis allows a further exploration of the microscopic interpretation of entropy.\cite{Arnetal2011}
As a matter of fact, it is worth stressing that after the equalization of pressures there are configurations in the vicinity of this mechanical equilibrium 
with greater global entropy, and the system will move towards these configurations.
As a consequence, the system will indeed access the different available microscopic configurations
and move as a result of a blind entropic process, in accordance with Boltzmann's basic ideas 
and his microscopic interpretation of entropy. The latter also furnishes an explanation on the additive property of entropy.

\newpage
\section*{Figures}

%




\end{document}